\documentclass[aip,jcp,amsmath,reprint,groupedaddress]{revtex4-1}
\pdfoutput=1
\usepackage[colorlinks]{hyperref}
\usepackage[utf8]{inputenc}
\usepackage{mathptmx}
\usepackage{graphicx}
\usepackage{microtype}

\newcommand{\vt}[1]{\mathbf{#1}}
\newcommand{\dd}{\mathrm{d}}

\begin{document}
\title{The Complemented System Approach: A Novel Method for Calculating the X-ray Scattering from Computer Simulations}
\author{Andrej Lajovic}
\author{Matija Tomšič}
\author{Andrej Jamnik}
\email{andrej.jamnik@fkkt.uni-lj.si}
\affiliation{Faculty of Chemistry and Chemical Technology, University of Ljubljana, Aškerčeva~5, SI-1000 Ljubljana, Slovenia}

\begin{abstract}
\advance\rightskip0.5in
In this paper, we review the main problem concerning the calculation of X-ray scattering of simulated model systems, namely their finite size. A novel method based on the Rayleigh–Debye–Gans approximation was derived, which allows sidestepping this issue by complementing the missing surroundings of each particle with an average image of the system. The method was designed to operate directly on particle configurations without an intermediate step (e.g., calculation of pair distribution functions): in this way, all information contained in the configurations was preserved. A comparison of the results against those of other known methods showed that the new method combined several favourable properties: an arbitrary $q$-scale, scattering curves free of truncation artifacts and good behaviour down to the theoretical lower limit of the $q$-scale. A test of computational efficiency was also performed to establish a relative scale between the speeds of all known methods: the reciprocal lattice approach, the brute force method, the Fourier transform approach and the newly presented complemented system approach.

\medskip
This article \href{http://link.aip.org/link/?jcp/133/174123}{appeared} in The Journal of Chemical Physics [doi:\href{http://dx.doi.org/10.1063/1.3502683}{10.1063/1.3502683}]

\copyright\ 2010 American Institute of Physics.
\end{abstract}


\maketitle

\section{Introduction}
The theoretical foundations of the theory of X-ray scattering were laid down nearly a century ago in the form of the Rayleigh–Debye–Gans approximation. \cite{art:rayleigh1914,art:debye1915,art:gans1925} Although more general theories were developed later, this approach still remains popular and is widely used to interpret and evaluate the results of various scattering experiments. The theory establishes a Fourier transform relation between the real space (the sample) and the reciprocal space (the scattering pattern). Moving between these two might therefore seem straightforward; however, in practice, going either way brings its own set of difficulties. \cite{book:saxs1982:glatter} This discussion focuses on the problem of calculating the scattering pattern of a model system constructed by computer simulation.

The main obstacle in such a calculation is the finiteness of the model system. The choice of the volume of the simulation cell is frequently rather limited—at the lower end, it is bounded by the maximum range of the correlations in the simulated system, while at the upper, the available CPU time is usually the limiting factor. At the first sight, one might be tempted to disregard the difference between the space equipped with periodic boundary conditions and the real (physical) space, and apply the Fourier transform to the raw particle configurations obtained from a simulation. However, such a na\"ive approach leads to severe truncation artifacts which completely engulf the relevant part of the scattering pattern and render the results unusable.

Various methods have been developed to overcome this problem: the most direct one is the reciprocal lattice approach, in which an infinite system is constructed by stacking simulation cells into a pseudo-crystal. \cite{art:frenkel1986} In consequence, the resulting scattering pattern consists of discrete scattering peaks whose positions depend only on the cell side length. While the results are otherwise favourable, such discreteness is clearly not a feature of the simulated system, but rather a consequence of the way the method deals with periodic boundary conditions.

Another method (the so-called ``brute force approach''), which we developed recently, applies the Debye equation to a set of polydisperse cubic cut-outs of the system and combines their scattering patterns in order to suppress the truncation artifacts. \cite{art:tomsic2007_jpcb} In contrast to the reciprocal lattice approach, this method yields scattering intensities at arbitrary angles and is thus more in accord with the true nature of the system. But while providing relatively satisfactory results, the numerical suppression of truncation artifacts is found to be less than optimal, leaving some remains visible in the form of small oscillations on the scattering curve, particularly at small angles. This is to be expected, since the method is based mainly on numerical grounds.

In the third—and perhaps most commonly used—\discretionary{}{}{}method, the structure factor for each pair of atom types is first determined by Fourier-transforming the corresponding pair distribution function. The total scattering is then calculated by multiplying these structure factors by appropriate particle form factors and summing up the results. \cite{art:smascm2002:pusey} Application of this method to the results of a computer simulation requires determination of the various pair distribution functions from the simulation data, preferably with good resolution and sufficiently small statistical uncertainty. This implies some sort of binning which inherently discards a part of the available information.

In the following discussion, we propose a method that applies the Debye equation directly to configuration snapshots in order to preserve all the information contained within them, while also compensating for the finite size of the simulation cell by complementing the system in a physically sensible way.

\vspace*{0pt plus 4 em}

\section{Theory}
The differential scattering cross-section per unit volume of a sample containing discrete particles is given by the Debye equation \cite{art:smascm2002:pusey}
\begin{equation}
\frac{\dd\Sigma}{\dd\Omega}(\vt{q}) = \frac{1}{V} \left\langle \sum_{j=1}^N \sum_{k=1}^N b_j(\vt{q}) b_k(\vt{q}) \exp\big[-i \vt{q} \cdot (\vt{r}_j - \vt{r}_k)\big] \right\rangle \,,
\label{eq:debye}
\end{equation}
where $\vt{q}$ represents the scattering vector, $V$ is the volume of the sample, $N$ is the number of particles in the system, indices~$j$ and~$k$ denote the $j$-th and the $k$-th particle, $b_j(\vt{q})$ and $b_k(\vt{q})$ are their scattering lengths and $\vt{r}_j$ and $\vt{r}_k$ their position vectors. The angle brackets denote the canonical average. Let us assume now that all the particles are spherical and that there are $T$ types of them; their corresponding numbers shall be designated by $N_1, N_2, \ldots, N_T$. The right-hand side of equation~(\ref{eq:debye}) can thus be rewritten as follows
\begin{multline}
\frac{1}{V} \Bigg\langle \iint\limits_V \sum_{A=1}^T \sum_{B=1}^T \sum_{j=1}^{N_A} \sum_{k=1}^{N_B} b_A(\vt{q}) b_B(\vt{q}) \delta(\vt{r'} - \vt{r}_{A:j}) \\
\times \delta(\vt{r''} - \vt{r}_{B:k}) \exp\big[-i \vt{q} \cdot (\vt{r'} - \vt{r''})\big] \dd\vt{r'} \dd\vt{r''} \Bigg\rangle \,.
\end{multline}
Integration is performed over the sample volume, the first two sums go over the atom types; $\delta(r)$ is the Dirac delta function, subscript $A{:}j$ means \textit{the $j$-th atom of type $A$} and subscript $B{:}k$ means \textit{the $k$-th atom of type $B$}.  For spherical particles, scattering lengths depend only on the atom type—since they are independent of the particles' orientation—and can be taken out of the integrals. Some further rearrangement results in
\begin{multline}
\frac{1}{V} \sum_{A=1}^T \sum_{B=1}^T b_A(\vt{q}) b_B(\vt{q}) \iint\limits_V \Bigg\langle \sum_{j=1}^{N_A} \sum_{k=1}^{N_B} \delta(\vt{r'} - \vt{r}_{A:j}) \\
\times \delta(\vt{r''} - \vt{r}_{B:k}) \Bigg\rangle \exp\big[-i \vt{q} \cdot (\vt{r'} - \vt{r''})\big] \dd\vt{r'} \dd\vt{r''} \,.
\label{eq:scat_int}
\end{multline}
The quantity in angle brackets can be transformed into~\cite{art:smascm2002:klein}
\begin{equation}
\delta(\vt{r'} - \vt{r''}) \rho^{(1)}_A(\vt{r'})\delta_{A,B} + \rho^{(2)}_{AB}(\vt{r'},\vt{r''}) \,,
\end{equation}
where $\rho^{(1)}_A(\vt{r'})$ and $\rho^{(2)}_{AB}(\vt{r'},\vt{r''})$ are the one- and two-particle number densities and $\delta_{A,B}$ is the Kronecker delta. Let us impose a further restriction, namely that the system be homogeneous. For such a case, one has
\begin{align}
\rho^{(1)}_A(\vt{r'}) &= \rho_A \\
\rho^{(2)}_{AB}(\vt{r'},\vt{r''}) &= \rho^{(2)}_{AB}(\vt{r'} - \vt{r''}) \,,
\end{align}
i.e., the one-particle density of type $A$ is equal to the average density of particles of type $A$, $\rho_A$, and the two-particle density of types $A$ and $B$ depends only on the difference between position vectors $\vt{r'}$ and $\vt{r''}$ (denoted below by~$\vt{r}$). One of the integrations in equation~(\ref{eq:scat_int}) can be therefore performed explicitly:
\begin{multline}
\iint\limits_V \left[ \delta(\vt{r'} - \vt{r''}) \rho^{(1)}_A(\vt{r'})\delta_{A,B} + \rho^{(2)}_{AB}(\vt{r'},\vt{r''}) \right] \\
\times\exp\big[-i \vt{q} \cdot (\vt{r'} - \vt{r''})\big] \dd\vt{r'} \dd\vt{r''}
\\
= V \left[ \rho_A\delta_{A,B} + \int_V \rho^{(2)}_{AB}(\vt{r}) \exp\big[-i\vt{q}\cdot\vt{r}\big] \dd\vt{r} \right] \,.
\label{eq:int_explicit}
\end{multline}
By introducing the pair correlation functions \cite{book:hansen_simple_liquids} $g^{(2)}_{AB}(\vt{r})$
\begin{equation}
g^{(2)}_{AB}(\vt{r}) = \frac{\rho^{(2)}_{AB}(\vt{r})}{\rho_A \rho_B} \,,
\end{equation}
equations (\ref{eq:scat_int}) and (\ref{eq:int_explicit}) can be combined into
\begin{multline}
\frac{\dd\Sigma}{\dd\Omega}(\vt{q}) = \sum_{A=1}^T \sum_{B=1}^T b_A(\vt{q}) b_B(\vt{q}) \\
\times \bigg[ \rho_A\delta_{A,B} + \rho_A \rho_B \int_V g^{(2)}_{AB}(\vt{r}) \exp[-i\vt{q}\cdot\vt{r}] \dd\vt{r} \bigg] \,.
\label{eq:vect_final}
\end{multline}

Frequently, the systems considered in X-ray scattering experiments are not only homogeneous, but also isotropic—meaning that both the functions in the above equation  and the result itself depend only on the length of the scattering vector $q = |\vt{q}|$. This allows one to perform rotational averaging on equations~(\ref{eq:debye}) and~(\ref{eq:vect_final}), yielding
\begin{equation}
\begin{split}
\frac{\dd\Sigma}{\dd\Omega}(q) &= \frac{1}{V} \left\langle \sum_{j=1}^N \sum_{k=1}^N b_j(q) b_k(q) \frac{\sin(q R_{jk})}{q R_{jk}} \right\rangle
\\
&= \sum_{A=1}^T \sum_{B=1}^T b_A(q) b_B(q) \bigg[\rho_A\delta_{A,B} \\
&\qquad + \rho_A\rho_B \int_0^\infty \kern-1ex g_{AB}(r) 4\pi r^2 \frac{\sin(qr)}{qr} \dd r \bigg] \,,
\end{split}
\label{eq:rotavg}
\end{equation}
where $R_{jk}$ is the distance between the $j$-th and $k$-th atom and $r = |\vt{r}|$. The right-hand side of this equation shows that one can calculate the scattering of a system if all pair distribution functions $g_{AB}(r)$ are known. This approach (called the \textit{Fourier transform approach} in this discussion) is especially viable if analytical solutions to $g_{AB}(r)$ can be found. A slightly modified version of the above equation is ordinarily used:
\begin{multline}
\frac{\dd\Sigma}{\dd\Omega}(q) = \sum_{A=1}^T \sum_{B=1}^T b_A(q) b_B(q) \Bigg[ \rho_A\delta_{A,B} \\
+ \rho_A\rho_B \bigg[ (2\pi)^3 \delta(q) + \int_0^\infty \kern-1ex \big( g_{AB}(r) - 1 \big) 4\pi r^2 \frac{\sin(qr)}{qr} \dd r \bigg] \Bigg] \,.
\label{eq:ft_usual}
\end{multline}
The term $(2\pi)^3 \delta(q)$ represents forward scattering, i.e., radiation scattered in the direction of the primary beam. Since this contribution cannot be measured, it is of little practical value and is thus frequently omitted from the result.

The Fourier transform approach can also be used with $g_{AB}(r)$ obtained from simulation data, provided that the resolution of these functions in the $r$-domain is high enough.

Equation (\ref{eq:rotavg}) reveals very clearly the origin of the unwanted truncation effects that arise when calculating the scattering from simulation data via the Debye equation; failing to take into account the terms with $R_{jk} > r_c$ (where $r_c$ is the cut-off distance, usually half of the simulation cell's side length) is equivalent to setting all pair distribution functions $g_{AB}(r)$ to zero for $r > r_c$. From the viewpoint of a particle in the centre of the cell, the density at large distances does not tend towards its average value within the system, but instead drops off sharply to zero when crossing the boundary of the cell. Unfortunately, a computer simulation does not provide any information about correlations over distances larger than $r_c$; however, a widely used criterion for selecting an appropriate simulation cell size is that over a distance $r_c = L/2$ ($L$ being the cell's side length), all interparticle correlations should vanish. A best guess would therefore be that at $r > r_c$, all pair correlation functions assume a constant value equal to their theoretical limiting value at large distances. A set of \textit{complemented} pair distribution functions $\gamma_{AB}(r)$ is then constructed:
\begin{equation}
\gamma_{AB}(r) = \begin{cases}
g^\mathrm{sim}_{AB}(r) & r \le r_c\\
\lim_{r \to\infty} g_{AB}(r) & r > r_c \,,
\end{cases}
\end{equation}
where $g^\mathrm{sim}_{AB}(r)$ are the pair distribution functions obtained from the simulation data. The limiting values of all pair distribution functions are $1$ by definition (strictly speaking, $\lim_{r \to\infty} g_{AB}(r)$ equals $1$ when $A\neq B$ and $1-1/N_A$ when $A=B$, but since we are considering the limiting value of $g_{AB}(r)$ in the thermodynamic sense, i.e., in an infinitely large system, $N_A$ grows over all limits and the term $1/N_A$ vanishes). Inserting $\gamma_{AB}(r)$ in place of $g_{AB}(r)$ in the right-hand side of equation (\ref{eq:rotavg}), we get
\begin{multline}
\frac{\dd\Sigma}{\dd\Omega} = \sum_{A=1}^T \sum_{B=1}^T b_A(q) b_B(q) \bigg[\rho_A\delta_{A,B} \\
+ \rho_A\rho_B \int_0^\infty \kern-1ex g^\mathrm{sim}_{AB}(r) 4\pi r^2 \frac{\sin(qr)}{qr} H(r_c - r) \dd r \bigg]
\\
+ \sum_{A=1}^T \sum_{B=1}^T b_A(q) b_B(q) \rho_A \rho_B \int_{r_c}^\infty 4\pi r^2 \frac{\sin(qr)}{qr} \dd  r \,.
\label{eq:part_gamma}
\end{multline}
$H(r)$ is the Heaviside step function. Note that the first term of the above equation differs from the right-hand side of equation~(\ref{eq:rotavg}) only in discarding distances larger than $r_c$. One could therefore get the same result by discarding all $R_{jk} > r_c$ in the left-hand side of equation~(\ref{eq:rotavg})
\begin{multline}
\sum_{A=1}^T \sum_{B=1}^T b_A(q) b_B(q) \bigg[\rho_A\delta_{A,B} \\
+ \rho_A\rho_B \int_0^\infty \kern-1ex g^\mathrm{sim}_{AB}(r) 4\pi r^2 \frac{\sin(qr)}{qr} H(r_c - r) \dd r \bigg] =
\\
= \frac{1}{V} \left\langle \sum_{j=1}^N \sum_{k=1}^N b_j(q) b_k(q) \frac{\sin(q R_{jk})}{q R_{jk}} H(r_c - R_{jk}) \right\rangle \,.
\label{eq:H_discarding_Rjk}
\end{multline}
The second term in equation (\ref{eq:part_gamma}) can be written as follows:
\begin{multline}
\sum_{A=1}^T \sum_{B=1}^T b_A(q) b_B(q) \rho_A \rho_B \int_{r_c}^\infty 4\pi r^2 \frac{\sin(qr)}{qr} \dd  r =
\\
= \left( \sum_{A=1}^T b_A(q) \rho_A \right)^2 \bigg( \int_0^\infty 4\pi r^2 \frac{\sin(qr)}{qr} \dd  r \\
- \int_0^{r_c} 4\pi r^2 \frac{\sin(qr)}{qr} \dd  r \bigg) \,.
\label{eq:boxterm_split}
\end{multline}
Although the first of the above integrals cannot be evaluated in the Riemann sense, it can be regarded as a distribution and is, in that respect, equivalent to $(2\pi)^3 \delta(q)$. Inserting equations (\ref{eq:H_discarding_Rjk}) and (\ref{eq:boxterm_split}) into equation (\ref{eq:part_gamma}), we get
\begin{multline}
\frac{\dd\Sigma}{\dd\Omega}(q) = \frac{1}{V} \left\langle \sum_{j=1}^N \sum_{k=1}^N b_j(q) b_k(q) \frac{\sin(q R_{jk})}{q R_{jk}} H(r_c - R_{jk}) \right\rangle
\\
+ \left( \sum_{A=1}^T b_A(q) \rho_A \right)^2 \bigg[ (2\pi)^3 \delta(q) \\
- \frac{4\pi}{q^3} \big[ \sin(qr_c) - qr_c \cos(qr_c) \big] \bigg] \,.
\label{eq:cs_final}
\end{multline}
The first term is the familiar Debye relation, but with only interparticle distances of less than $r_c$ taken into account: this term represents the scattering of the explicit part of the system (i.e., the simulated particles). This contribution is complemented by the second term which arises from light interference between the explicit part and the averaged surroundings. A real-world example of these two contributions is shown later in the text in figure \ref{fig:terms}. The term $(2\pi)^3 \delta(q)$ again represents the forward scattering, which is usually of no interest and can be omitted altogether at $q > 0$.

\section{Numerical tests}
In order to evaluate the characteristics of the complemented system approach compared to other aforementioned methods, a test run was conducted using the simulation data from our previous study on aldehydes. \cite{art:lajovic2009} The TraPPE-UA (Transferable Potential for Phase Equilibria—United Atom) force field \cite{art:stubbs2004} was used to model the aldehyde. In this model, CH$_3$, CH$_2$ and CH groups are treated as single sites: an aldehyde molecule therefore consists of a linear chain of united CH$_{\rm x}$ atoms with an oxygen atom bonded to the terminal one.

The simulation run comprised 300 pentanal molecules at 25\,$^\circ$C in a simulation box with a side length of 37.7025~\AA. The configurational bias Monte Carlo technique was employed. First, the system was equilibrated for 10\,000 cycles (one cycle consisted of 300 Monte Carlo steps); after that, a production run of 40\,000 cycles followed. During this run, a snapshot of the configuration was saved each 400 steps; a total of 100 independent configuration snapshots were obtained this way.

X-ray scattering curves were calculated directly from the simulation data by each respective method. In the Fourier transform method, the pair distribution functions were estimated by binning the interparticle distances. To evaluate the impact of the bin width on the results, four different bin widths were used: $0.5$~\AA, $0.2$~\AA, $0.1$~\AA\ and $0.05$~\AA.

The computational efficiency of the methods was assessed by measuring the time needed to process a set of 100 configurations, each of them comprising $N$ particles. A series of such sets was generated, with $N$ varying between one hundred and several thousand. For the Fourier transform (FT) approach, the brute force (BF) method and the complemented system (CS) approach, a $q$-range of $0$~\AA$^{-1}$ to $2.5$~\AA$^{-1}$ in steps of $0.05$~\AA$^{-1}$ was used (totalling 50 points on the scattering curve), while for the reciprocal lattice (RL) approach, the Miller indices were limited to $h,k,l \le 10$, leading to 85 points within a $q$-range of $0.17$~\AA$^{-1}$ to $1.7$~\AA$^{-1}$. Pair distribution functions for use in FT were collected with a bin width of $0.1$~\AA. The algorithms were implemented as follows: BF as a FORTRAN program, RL and CS as a C program and FT as a GNU Octave script.

\section{Results and discussion}
\begin{figure}[t]
{\centering\input{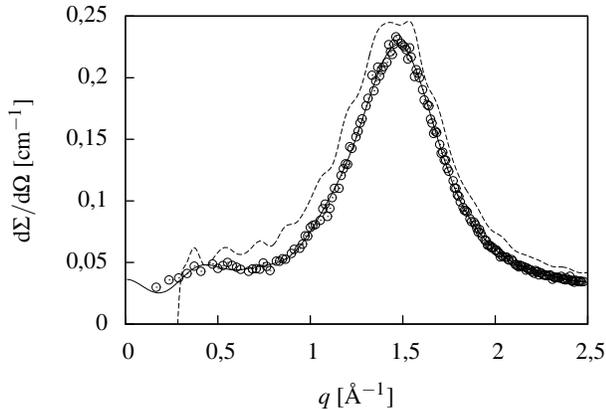}}
\caption{\label{fig:comparison}Comparison of results: the scattering curve of pentanal calculated via different methods: the reciprocal lattice approach (circles), the brute force method (dashed line) and the complemented system approach (solid line).}
\end{figure}
\begin{figure}[t]
{\centering\input{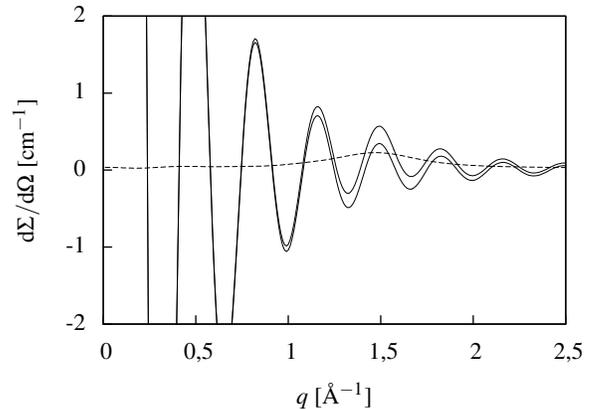}}
\caption{\label{fig:terms}The two contributions of equation (\ref{eq:cs_final}). The first term produces the upper solid curve and the second term produces the lower solid curve; the latter is plotted with a negative sign in order to facilitate the comparison. Subtracting the lower curve from the upper yields the scattering curve corrected for the finite-size effects; it is shown here in a dashed line.}
\end{figure}

The results of the reciprocal lattice approach, the brute force method and the complemented system approach are shown in figure~\ref{fig:comparison}. It is evident that the result of the CS almost completely matches the result of the RL with the only discernible difference arising at very low angles ($q$ less than approximately $0.3$~\AA$^{-1}$). It must be taken into account, though, that the simulation reproduces only correlations over distances below one-half of the simulation cell's side length. (In fact, distances between $L/2$ and $L\sqrt{3}/2$, $L$ being the simulation cell's side length, are also reproduced, but with lesser accuracy. Nevertheless, one must keep in mind that the CS method, at least in the present implementation, explicitly ignores distances longer than $L/2$.) Any scattering curve calculated from simulation data is therefore valid only down to $q_\mathrm{min}=4\pi/L$, which for the curves shown equals $0.33$~\AA$^{-1}$. The observed behaviour of the scattering curves in figure~\ref{fig:comparison} is in good agreement with this value, exemplifying the fact that differences in the results below $q_\mathrm{min}$ are actually to be expected due to the different methods used. We can conclude that the RL and CS give almost identical results, which affirms the conceptual consistency of the two approaches.

The BF, on the other hand, gives a somewhat higher intensity with slight superimposed oscillations which increase in amplitude, particularly at lower $q$-values. Both characteristics were observed to be regular features of this method.

According to equation (\ref{eq:cs_final}), the scattering is composed of two additive contributions: one due to the explicit part of the system and the other due to the averaged surroundings. To illustrate the high importance of the latter, both contributions are plotted separately in figure~\ref{fig:terms}, together with the resulting scattering curve. One can see that the finite-size effects are drastic—the maximum value of the first term is about three orders of magnitude larger than the peak value of the final scattering curve. For finite-size simulated systems, the Debye equation is clearly not usable without an additional correcting term.

\begin{figure}[tb]
{\centering\input{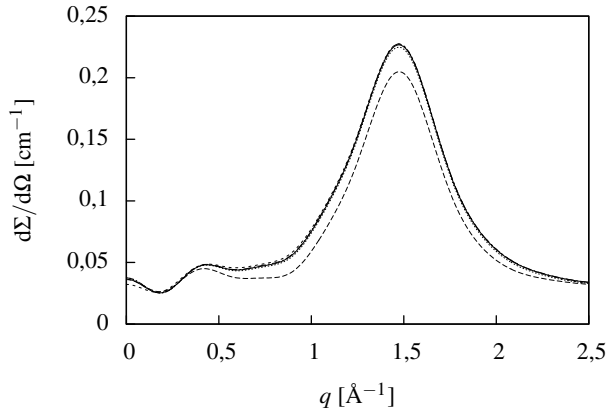}}
\caption{\label{fig:cs+g}The scattering curves of pentanal calculated via the complemented system approach (solid line) and via Fourier-transformation of the pair distribution functions. Several series of pair distribution functions were generated, each with a different binning width: $0.5$~\AA\ (long dashed), $0.2$~\AA\ (short dashed), $0.1$~\AA\ (dotted) and $0.05$~\AA\ (dot-dashed). See also figure~\ref{fig:cs+g_closeup} for a close up.}
\end{figure}
\begin{figure}[tb]
{\centering\input{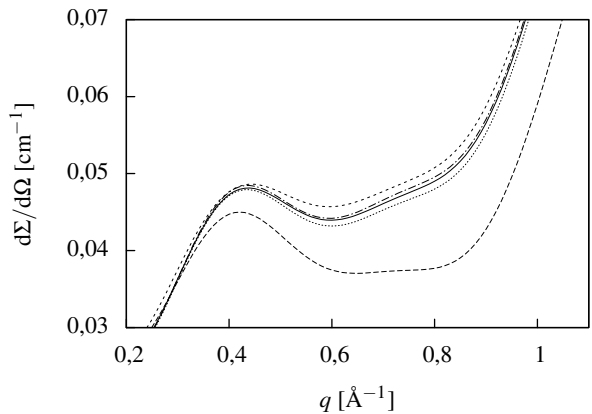}}
\caption{\label{fig:cs+g_closeup}Close up of figure~\ref{fig:cs+g}. Bin widths used: $0.5$~\AA\ (long dashed), $0.2$~\AA\ (short dashed), $0.1$~\AA\ (dotted) and $0.05$~\AA\ (dot-dashed). The result of the complemented system approach is shown as a solid line.}
\end{figure}

Another set of results, this time focusing on the method based on Fourier transforms of the pair distribution functions, is shown in figure~\ref{fig:cs+g} together with the results of the CS. For each FT scattering curve, a different bin width was used when constructing the pair distribution functions, in order to assess its effects on the final result. Excepting the case of the largest bin width, the curves are very similar and a close up is shown in figure~\ref{fig:cs+g_closeup} to reveal the differences. As expected, the results of the FT method approach the results of the CS as the bin width decreases (these two methods theoretically become identical in the limit when bin width approaches zero). However, it is interesting to note that this asymptotic approach does not proceed smoothly from one side; instead the results of the FT fluctuate around the limiting value with an ever decreasing amplitude.

This observation reveals a plausible use for the CS even in cases where many similar calculations must be made quickly, which usually makes the FT the preferred method due to its speed. The CS can be viewed as an optimal case of the FT (bin width infinitely small) and its results can therefore be used to gauge the effect of bin width in order to select an acceptable value in a particular context. When speed is not the limiting factor, the CS can be used directly to get an optimal result immediately.

\begin{figure}[t]
{\centering\input{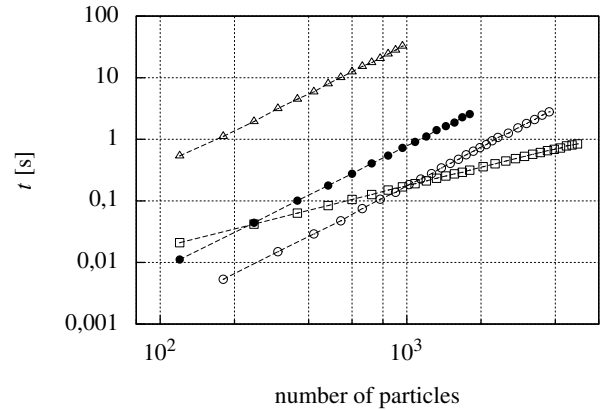}}
\caption{\label{fig:scaling}Computational efficiency of the methods. The diagram shows the time required to process a configuration with a certain number of particles. Symbols denote: RL (squares), BF (triangles), FT (empty circles), CS (filled circles).}
\end{figure}
Indeed, the computational efficiency is usually an important consideration in selection of a method. To provide a comparison between the efficiencies of the methods used, a series of test runs was conducted solely for that purpose. Even though the actual run times of the programs might not be directly applicable in general due to their dependence on the implementation, the compiler/interpreter and the underlying hardware, we feel that on a relative scale, the results are telling enough.

The relation between the number of particles in a configuration, $N$, and the time needed to process it is shown in figure~\ref{fig:scaling}. An immediately apparent fact is that the time complexity of the RL method is $O(N)$ in contrast to the other three methods which share a $O(N^2)$ time complexity. Thus, for a large enough number of particles, the RL method will always be the most efficient—at a price of having fixed $q$-values and exhibiting somewhat larger statistical uncertainties in the resulting curves. Amongst the other methods, FT is the fastest, followed by CS (approximately four times slower) and BF (approximately 200 times slower than FT).

It should be stressed that the majority of the time needed in an FT calculation is spent gathering the pair distribution functions; the calculation of scattering actually takes a negligible amount of time which is also independent of~$N$. Also, the binning width and the density of $q$-points do not affect the run times appreciably; they do affect the memory usage, though.

\section{Conclusion}
In this paper, we presented a novel method for calculating the X-ray scattering of a (simulated) model system. It operates directly on particle configurations and thus avoids the need to calculate the pair distribution functions: in this way, all the information contained in the configuration is preserved. Correlations over distances larger than some cut-off distance (in the usual case of a computer simulation, half of the cell's side length) are handled in a best-effort way by setting the pair distribution functions to their theoretical limiting value: such an approximation effectively ``masks'' the finiteness of the system by complementing the missing surroundings of each particle with an average image of the system. We showed that this approach can in fact be regarded as a special case of the widely used Fourier transform method, giving directly the results that the latter method would produce in the limit of an infinitely small bin width. Numerical tests showed that the results also compare favourably to the results of other known methods.

\begin{acknowledgments}
We dedicate this work to Professor Otto Glatter (University of Graz, Austria) on the occasion of his 65$^{\rm th}$ birthday. We are thankful to him for many years of fruitful scientific cooperation and for the kind hospitality that we experienced during our visits to his group.

We acknowledge financial support from the Slovenian Research Agency through the Physical Chemistry Research Programme 0103-0201 and from the Federal ministry for education, science, and culture of Austria (BI-AT/09-10-022).
\end{acknowledgments}

\bibliography{csappr}

\begin{thebibliography}{11}%
\makeatletter
\providecommand \@ifxundefined [1]{%
 \@ifx{#1\undefined}
}%
\providecommand \@ifnum [1]{%
 \ifnum #1\expandafter \@firstoftwo
 \else \expandafter \@secondoftwo
 \fi
}%
\providecommand \@ifx [1]{%
 \ifx #1\expandafter \@firstoftwo
 \else \expandafter \@secondoftwo
 \fi
}%
\providecommand \natexlab [1]{#1}%
\providecommand \enquote  [1]{``#1''}%
\providecommand \bibnamefont  [1]{#1}%
\providecommand \bibfnamefont [1]{#1}%
\providecommand \citenamefont [1]{#1}%
\providecommand \href@noop [0]{\@secondoftwo}%
\providecommand \href [0]{\begingroup \@sanitize@url \@href}%
\providecommand \@href[1]{\@@startlink{#1}\@@href}%
\providecommand \@@href[1]{\endgroup#1\@@endlink}%
\providecommand \@sanitize@url [0]{\catcode `\\12\catcode `\$12\catcode
  `\&12\catcode `\#12\catcode `\^12\catcode `\_12\catcode `\%12\relax}%
\providecommand \@@startlink[1]{}%
\providecommand \@@endlink[0]{}%
\providecommand \url  [0]{\begingroup\@sanitize@url \@url }%
\providecommand \@url [1]{\endgroup\@href {#1}{\urlprefix }}%
\providecommand \urlprefix  [0]{URL }%
\providecommand \Eprint [0]{\href }%
\providecommand \doibase [0]{http://dx.doi.org/}%
\providecommand \selectlanguage [0]{\@gobble}%
\providecommand \bibinfo  [0]{\@secondoftwo}%
\providecommand \bibfield  [0]{\@secondoftwo}%
\providecommand \translation [1]{[#1]}%
\providecommand \BibitemOpen [0]{}%
\providecommand \bibitemStop [0]{}%
\providecommand \bibitemNoStop [0]{.\EOS\space}%
\providecommand \EOS [0]{\spacefactor3000\relax}%
\providecommand \BibitemShut  [1]{\csname bibitem#1\endcsname}%
\let\auto@bib@innerbib\@empty
\bibitem [{\citenamefont {Rayleigh}(1914)}]{art:rayleigh1914}%
  \BibitemOpen
  \bibfield  {author} {\bibinfo {author} {\bibfnamefont {D.~W.}\ \bibnamefont
  {Rayleigh}},\ }\href@noop {} {\bibfield  {journal} {\bibinfo  {journal}
  {Proc. R. Soc. London Ser. A}\ }\textbf {\bibinfo {volume} {90}},\ \bibinfo
  {pages} {219} (\bibinfo {year} {1914})}\BibitemShut {NoStop}%
\bibitem [{\citenamefont {Debye}(1915)}]{art:debye1915}%
  \BibitemOpen
  \bibfield  {author} {\bibinfo {author} {\bibfnamefont {P.}~\bibnamefont
  {Debye}},\ }\href@noop {} {\bibfield  {journal} {\bibinfo  {journal} {Ann.
  Phys.}\ }\textbf {\bibinfo {volume} {46}},\ \bibinfo {pages} {809} (\bibinfo
  {year} {1915})}\BibitemShut {NoStop}%
\bibitem [{\citenamefont {Gans}(1925)}]{art:gans1925}%
  \BibitemOpen
  \bibfield  {author} {\bibinfo {author} {\bibfnamefont {R.}~\bibnamefont
  {Gans}},\ }\href@noop {} {\bibfield  {journal} {\bibinfo  {journal} {Ann.
  Phys.}\ }\textbf {\bibinfo {volume} {76}},\ \bibinfo {pages} {29} (\bibinfo
  {year} {1925})}\BibitemShut {NoStop}%
\bibitem [{\citenamefont {Glatter}(1982)}]{book:saxs1982:glatter}%
  \BibitemOpen
  \bibfield  {author} {\bibinfo {author} {\bibfnamefont {O.}~\bibnamefont
  {Glatter}},\ }in\ \href@noop {} {\emph {\bibinfo {booktitle} {Small Angle
  {X}-ray Scattering}}},\ \bibinfo {editor} {edited by\ \bibinfo {editor}
  {\bibfnamefont {O.}~\bibnamefont {Glatter}}\ and\ \bibinfo {editor}
  {\bibfnamefont {O.}~\bibnamefont {Kratky}}}\ (\bibinfo  {publisher} {Academic
  press inc.},\ \bibinfo {address} {London},\ \bibinfo {year} {1982})\ pp.\
  \bibinfo {pages} {119--196}\BibitemShut {NoStop}%
\bibitem [{\citenamefont {Frenkel}\ \emph {et~al.}(1986)\citenamefont
  {Frenkel}, \citenamefont {Vos}, \citenamefont {de~Kruif},\ and\ \citenamefont
  {Vrij}}]{art:frenkel1986}%
  \BibitemOpen
  \bibfield  {author} {\bibinfo {author} {\bibfnamefont {D.}~\bibnamefont
  {Frenkel}}, \bibinfo {author} {\bibfnamefont {R.~J.}\ \bibnamefont {Vos}},
  \bibinfo {author} {\bibfnamefont {C.~G.}\ \bibnamefont {de~Kruif}}, \ and\
  \bibinfo {author} {\bibfnamefont {A.}~\bibnamefont {Vrij}},\ }\href@noop {}
  {\bibfield  {journal} {\bibinfo  {journal} {The Journal of Chemical Physics}\
  }\textbf {\bibinfo {volume} {84}},\ \bibinfo {pages} {4625} (\bibinfo {year}
  {1986})}\BibitemShut {NoStop}%
\bibitem [{\citenamefont {Tom\v{s}i\v{c}}\ \emph {et~al.}(2007)\citenamefont
  {Tom\v{s}i\v{c}}, \citenamefont {Jamnik}, \citenamefont {Fritz-Popovski},
  \citenamefont {Glatter},\ and\ \citenamefont
  {Vl\v{c}ek}}]{art:tomsic2007_jpcb}%
  \BibitemOpen
  \bibfield  {author} {\bibinfo {author} {\bibfnamefont {M.}~\bibnamefont
  {Tom\v{s}i\v{c}}}, \bibinfo {author} {\bibfnamefont {A.}~\bibnamefont
  {Jamnik}}, \bibinfo {author} {\bibfnamefont {G.}~\bibnamefont
  {Fritz-Popovski}}, \bibinfo {author} {\bibfnamefont {O.}~\bibnamefont
  {Glatter}}, \ and\ \bibinfo {author} {\bibfnamefont {L.}~\bibnamefont
  {Vl\v{c}ek}},\ }\href@noop {} {\bibfield  {journal} {\bibinfo  {journal}
  {Journal of Physical Chemistry B}\ }\textbf {\bibinfo {volume} {111}},\
  \bibinfo {pages} {1738} (\bibinfo {year} {2007})}\BibitemShut {NoStop}%
\bibitem [{\citenamefont {Pusey}(2002)}]{art:smascm2002:pusey}%
  \BibitemOpen
  \bibfield  {author} {\bibinfo {author} {\bibfnamefont {P.~N.}\ \bibnamefont
  {Pusey}},\ }in\ \href@noop {} {\emph {\bibinfo {booktitle} {Neutrons,
  {X}-Rays and Light: Scattering Methods Applied to Soft Condensed Matter}}},\
  \bibinfo {series and number} {North-Holland Delta Series},\ \bibinfo {editor}
  {edited by\ \bibinfo {editor} {\bibfnamefont {P.}~\bibnamefont {Lindner}}\
  and\ \bibinfo {editor} {\bibfnamefont {T.}~\bibnamefont {Zemb}}}\ (\bibinfo
  {publisher} {Elsevier Science B.V.},\ \bibinfo {address} {Amsterdam},\
  \bibinfo {year} {2002})\ pp.\ \bibinfo {pages} {3--21}\BibitemShut {NoStop}%
\bibitem [{\citenamefont {Klein}(2002)}]{art:smascm2002:klein}%
  \BibitemOpen
  \bibfield  {author} {\bibinfo {author} {\bibfnamefont {R.}~\bibnamefont
  {Klein}},\ }in\ \href@noop {} {\emph {\bibinfo {booktitle} {Neutrons,
  {X}-Rays and Light: Scattering Methods Applied to Soft Condensed Matter}}},\
  \bibinfo {series and number} {North-Holland Delta Series},\ \bibinfo {editor}
  {edited by\ \bibinfo {editor} {\bibfnamefont {P.}~\bibnamefont {Lindner}}\
  and\ \bibinfo {editor} {\bibfnamefont {T.}~\bibnamefont {Zemb}}}\ (\bibinfo
  {publisher} {Elsevier Science B.V.},\ \bibinfo {address} {Amsterdam},\
  \bibinfo {year} {2002})\ pp.\ \bibinfo {pages} {351--379}\BibitemShut
  {NoStop}%
\bibitem [{\citenamefont {Hansen}\ and\ \citenamefont
  {McDonald}(1986)}]{book:hansen_simple_liquids}%
  \BibitemOpen
  \bibfield  {author} {\bibinfo {author} {\bibfnamefont {J.~P.}\ \bibnamefont
  {Hansen}}\ and\ \bibinfo {author} {\bibfnamefont {I.~R.}\ \bibnamefont
  {McDonald}},\ }\href@noop {} {\emph {\bibinfo {title} {Theory of simple
  liquids}}}\ (\bibinfo  {publisher} {Academic press, Inc.},\ \bibinfo
  {address} {London},\ \bibinfo {year} {1986})\BibitemShut {NoStop}%
\bibitem [{\citenamefont {Lajovic}\ \emph {et~al.}(2009)\citenamefont
  {Lajovic}, \citenamefont {Tom\v{s}i\v{c}}, \citenamefont {Fritz-Popovski},
  \citenamefont {Vl\v{c}ek},\ and\ \citenamefont {Jamnik}}]{art:lajovic2009}%
  \BibitemOpen
  \bibfield  {author} {\bibinfo {author} {\bibfnamefont {A.}~\bibnamefont
  {Lajovic}}, \bibinfo {author} {\bibfnamefont {M.}~\bibnamefont
  {Tom\v{s}i\v{c}}}, \bibinfo {author} {\bibfnamefont {G.}~\bibnamefont
  {Fritz-Popovski}}, \bibinfo {author} {\bibfnamefont {L.}~\bibnamefont
  {Vl\v{c}ek}}, \ and\ \bibinfo {author} {\bibfnamefont {A.}~\bibnamefont
  {Jamnik}},\ }\href@noop {} {\bibfield  {journal} {\bibinfo  {journal}
  {Journal of Physical Chemistry B}\ }\textbf {\bibinfo {volume} {113}},\
  \bibinfo {pages} {9429} (\bibinfo {year} {2009})}\BibitemShut {NoStop}%
\bibitem [{\citenamefont {Stubbs}, \citenamefont {Potoff},\ and\ \citenamefont
  {Siepmann}(2004)}]{art:stubbs2004}%
  \BibitemOpen
  \bibfield  {author} {\bibinfo {author} {\bibfnamefont {J.~M.}\ \bibnamefont
  {Stubbs}}, \bibinfo {author} {\bibfnamefont {J.~J.}\ \bibnamefont {Potoff}},
  \ and\ \bibinfo {author} {\bibfnamefont {J.~I.}\ \bibnamefont {Siepmann}},\
  }\href@noop {} {\bibfield  {journal} {\bibinfo  {journal} {J. Phys. Chem. B}\
  }\textbf {\bibinfo {volume} {108}},\ \bibinfo {pages} {17596} (\bibinfo
  {year} {2004})}\BibitemShut {NoStop}%
\end{thebibliography}%

\end{document}